\RequirePackage{rotating}
\documentclass{jfm}
\usepackage{textcomp}
\usepackage{graphicx}
\usepackage{float}
\usepackage{natbib}
\usepackage{epsfig}
\usepackage[dvipsnames]{xcolor}
\usepackage{framed}
\usepackage{lipsum}
\graphicspath{{figures/}}
\usepackage{color}
\usepackage{soul}
\usepackage{caption}
\usepackage{rotating}
\usepackage{ulem}
\usepackage{tikz}
\usepackage{bm}
\usepackage{subcaption}
\usepackage{stackengine}
\usepackage{amsmath}
\usepackage{mathtools}
\usepackage{calc}
\usepackage{accents}
\newcommand{\dbtilde}[1]{\accentset{\approx}{#1}}

\begin{document}
\newtheorem{lemma}{Lemma}
\newtheorem{corollary}{Corollary}
\newcommand{\comment}[1]{}
\newcommand*\mean[1]{\bar{#1}}
\newcommand{\red}[1]{\textcolor{red}{#1}}
\newcommand{\blue}[1]{\textcolor{blue}{#1}}

\DeclareRobustCommand\full  {\tikz[baseline=-0.6ex]\draw[thick] (0,0)--(0.5,0);}
\DeclareRobustCommand\dotted{\tikz[baseline=-0.6ex]\draw[thick,dotted] (0,0)--(0.54,0);}
\DeclareRobustCommand\dashed{\tikz[baseline=-0.6ex]\draw[thick,dashed] (0,0)--(0.54,0);}
\DeclareRobustCommand\chain {\tikz[baseline=-0.6ex]\draw[thick,dash dot dot] (0,0)--(0.5,0);}
\newcommand{\circlemarker}{\raisebox{0.5pt}{\tikz{\node[draw,scale=0.65,circle,fill=none](){};}}}
\newcommand{\squaremarker}{\raisebox{0.5pt}{\tikz{\node[draw,scale=0.8,rectangle,fill=none](){};}}}
\newcommand{\diamondmarker}{\raisebox{0pt}{\tikz{\node[draw,scale=0.65,rectangle,fill=none,rotate=45](){};}}}
\newcommand{\trianglemarker}{\raisebox{0pt}{\tikz{\node[draw,scale=0.65,(0,0)--(2,0)--(1,1)--cycle,fill=none](){};}}}

\shorttitle{Surface Pressure Spectra and Vorticity in a Turbulent Boundary Layer} 
\shortauthor{S. Glegg, S. Verma, L. Denissova} 
\title{The Relationship Between Surface Pressure Spectra and Vorticity in a Turbulent Boundary Layer}
\author
 {
Stewart Glegg\aff{1},
Siddhartha Verma\aff{1,2}, and
 Lyubov Denissova\aff{1}\corresp{\email{sglegg@fau.edu}}
  }
\affiliation
{
\aff{1}
Department of Ocean and Mechanical Engineering, Florida Atlantic University, Boca Raton, FL 33431, USA
\aff{2}
Harbor Branch Oceanographic Institute, Florida Atlantic University, Fort Pierce, FL 34946, USA
}
\maketitle
\begin{abstract}
The modeling of surface pressure wave number spectra beneath a turbulent boundary layer is reviewed and reconsidered in terms of the vorticity in the flow. Using a solution based on the vorticity equation and Squires theorem, which was originally given by Chase(1991), it is shown that the complete solution for surface pressure spectrum can be specified using the non-linear turbulence-turbulence interaction terms as sources. It is then shown that the surface pressure can be directly related to the vorticity in the flow. The results are checked against a Direct Numerical Simulation of a channel flow configuration. It is shown that the vorticity associated with the wall shear stress decays rapidly over a distance of about seven wall units and so the outer vorticity dominates in the majority of the flow. The vorticity correlation functions are evaluated and it is shown that the length scale of the vorticity normal the wall is only about ten wall units so the flow can be modeled as the superposition of uncorrelated vortex sheets. Finally a model for the surface pressure wave number spectrum is developed by modeling the vorticity spectrum based on the distribution of root mean square (rms) vorticity in the flow, and an integral length scale that is about ten wall units. Using inputs from the numerical simulation, a good fit to existing empirical models for surface pressure spectra is obtained.
\end{abstract}

\section{Introduction}
\label{sec:introduction}
The surface pressure fluctuations beneath a fully developed turbulent boundary layer are important in many engineering applications. Surface pressure perturbations can be the source of panel vibration on aircraft and marine vehicles \citep{Graham1997,Blake2017} and, using Amiet’s theory \citep{Amiet1976}, the cause of  trailing edge noise which is one of the most important noise sources on aircraft airframes, fans, wind turbines and helicopter rotors. For these reasons there has been considerable attention given to the prediction of surface pressure spectra and wavenumber spectra, and how they are related to the characteristics of the turbulent boundary layer (see \citet{Blake2017} for a recent review).

The approach to modeling the surface pressure spectrum below a turbulent boundary layer has been based on the reduction of experimental measurements to obtain empirical formulas \citep{Graham1997,Chase1991,Goody2004}, the use of Direct Numerical Simulations (DNS) of a channel flow \citep{Chang1999,Kim1989,Anantharamu2020}, Large Eddy Simulations (LES) over rough walls \citep{Ji2012}, or analytical modeling based on the solution to a Poisson's equation \citep{Kraichnan1956} or Lighthill’s Acoustic Analogy \citep{Ffowcs-Williams1965}. The modeling based on empirical formulas is limited to the types of flow in which the measurements were made, which are almost exclusively zero pressure gradient turbulent flows over smooth or rough walls. Similarly the DNS and LES calculations have been limited to simple flows, and are challenging in terms the computational storage requirements to ensure the convergence of time and space averages. None of these approaches are able to identify the flow characteristics that are responsible for the pressure perturbations in a less restricted flow environment, such as the fully three dimensional flow near the trailing edge or tip of a rotating fan blade. 

To extend the empirical and computational models to a more complex flow type, analytical models of the flow characteristics that cause the surface pressure fluctuations are required. In general, most of the modeling to date has been based on the use of Lighthill's Acoustic Analogy, or a Poisson's equation that is the same as Lighthill's equation in the incompressible limit. These models have led to the development of the Blake-TNO model \citep{Blake2017, Grasso2019} which specify the surface pressure based on two source terms that are derived from Lighthill's stress tensor, $T_{ij}$. The first of these terms is the so called fast term that depends on the mean velocity gradient normal to the wall and the unsteady velocity component normal to the wall. This term is linear in perturbation quantities, and most readily modeled using homogeneous turbulence wavenumber spectrum models and an assumed uniform convection speed \citep{Blake2017}. The second source term is based on the turbulence-turbulence interactions which cause a non-linear quantity, sometimes referred to as the slow term. The relative magnitude of these two terms is hard to assess because the non-linear interactions are almost impossible to measure or model, and can only be realistically obtained from numerical approaches such as LES or DNS. Many studies \citep{Grasso2019} have argued that the linear term dominates in certain frequency ranges and have modeled the surface pressure spectra based on the linear term alone. This suggests that the surface pressure spectra are determined by the mean flow gradient and the vertical component of the unsteady velocity fluctuations. Then by using a suitable model for the wavenumber spectra of the turbulence in the boundary layer, combined with a number of assumptions of the turbulence lengthscales, and the rms turbulence velocity distribution obtained from RANS calculations, a surface pressure spectrum model is obtained that is sensitive to modifications in the mean flow.

The problem with this approach is that it assumes that the turbulent velocity fluctuations are known precisely throughout the flow. If there is a DNS calculation of the flow then the source terms for the Poisson's equation can be evaluated exactly and the distribution of `apparent' or modelled source strength defined. This was recently the focus of the study by \citet{Anantharamu2020}, who used a DNS of a channel flow to define the terms on the right side of the Poisson's equation, and a Greens function to obtain the apparent distribution of sources that caused the surface pressure fluctuations. We use the term `apparent' because, as was shown by \citet{Hariri1985}, the pressure fluctuations throughout the flow and the local normal velocity are coupled, so one cannot be treated as a source for the other. In their study of wall shear stress, \citet{Hariri1985} solved both the Poisson's equation for the pressure and the vertical component of the momentum equation as two coupled partial differential equation with two unknowns. The right side of these two equations were non-linear, turbulence-turbulence interaction terms. It is often argued that this is a complete solution of the Navier Stokes equations because, in the wavenumber frequency domain, the non-linear terms are uncoupled with the linear terms and so may be calculated separately as `true' source terms. 

Separating the linear terms in the Navier Stokes equations from the non-linear terms for a parallel boundary layer flow to obtain the surface pressure was first carried out by \citet{Landahl1967} using the inhomogeneous form of the Orr-Sommerfeld equation to solve for the vertical velocity fluctuations in terms of non-linear source terms. The surface pressure was then obtained by integrating the wall normal component of the momentum equation. A modal solution to the Orr-Sommerfeld equations was used and the surface pressure obtained as a modal expansion. All the modes were found to decay with downstream distance and, for a homogeneous flow, the dominant modes were those with phase speeds with the smallest imaginary parts. A similar approach was used by \citet{Chase1991} who re-cast the vorticity equation using Squires transformation to obtain two uncoupled equations for the unsteady velocity in a parallel shear flow. The non-linear terms were defined as source terms in the frequency wavenumber domain, and the surface pressure and wall shear stress specified directly from the momentum equations evaluated at the wall. The source terms for the pressure fluctuations was then specified in terms of only three components of the fluctuating Reynolds shear stress. \citet{Chase1991} solved the Orr-Sommerfeld equation in the limit that the streamwise wavenumber tended to zero, and defined the vertical velocity perturbations and the surface pressure in terms of the non-linear source terms by using the method of variation of parameters.  

In all these approaches the most difficult part of the modeling is to correctly specify the non-linear source terms. Without recourse to LES or DNS there are no empirical or semi-empirical models for these terms, or well accepted insights as how they are a function of the mean flow and its statistics. In this paper we will address this problem by simplifying Chase’s (1991) analysis so that the source term is limited to the unsteady `outer' vorticity in the flow. We will define this as the vorticity in the flow with the vorticity caused by the wall shear stress subtracted. We will show analytically, and by using DNS calculations, that the surface pressure spectrum can be specified as an integral of the outer vorticity with an exponentially decaying weighting factor. This has some important implications for modeling more complex flows such as boundary layers with non-zero pressure gradients, rough wall boundary layers and boundary layers downstream of obstructions. In those flows modeling the convected vorticity is a tractable problem. For many practical problems it may also have a local solution when the scale of the vorticity is sufficiently small that it can be embedded into a mean flow that changes on a much larger scale.

The layout of this paper is as follows. In section 2 we will discuss the approach to surface pressure based on Lighthill's Analogy and the Poisson's equation, specifically highlighting the role of viscous shear stress at the wall. In section 3 we will specify the Poisson's equation for a parallel shear flow in the wavenumber frequency domain and show how there are two coupled equations that must be solved for both the fluctuating pressure and the vertical velocity component. In section 4 we will use the vorticity equation and Squires transformation to derive the inhomogeneous Orr-Sommerfeld equation as was done by \citet{Chase1991}, and relate the surface pressure to the viscous shear stress at the wall. We then specify the inner and outer vorticity and show how the surface pressure can be specified in terms of the outer vorticity alone. In section 5 we use a DNS calculation for a channel flow to verify the analytical results given in section 4. In section 6 we specify the wavenumber spectrum of the surface pressure based on simple model of the vorticity fluctuations in a turbulent boundary layer, and compare the predicted spectrum to the Goody model for surface pressure spectra. In section 7 details are given of the vorticity correlation functions obtained from the DNS data, and it is shown that they are consistent with the simple model of the vorticty spectrum used in section 6.

\section{The Sources of Surface Pressure Fluctuations based on Poisson's Equation}
\label{sec:sourcePressure}

It is natural \citep{Ffowcs-Williams1965} to start a derivation of the pressure fluctuations in a flow by considering Lighthill's Acoustic Analogy which describes the propagation of acoustic waves in a stationary medium that are generated by a region of turbulence. In a stationary medium the density perturbations are given by the inhomogeneous wave equation:
\begin{equation} 
\dfrac{\partial^2 \rho'}{\partial t^2} -  c_\infty^2 \dfrac{\partial^2 \rho'}{\partial x_i^2} = \dfrac{\partial^2 T_{ij}}{\partial x_i \partial x_j}
\end{equation}
and provided that the density is only a function of pressure and the speed of sound is constant, as is the case in most flows of interest at low Mach number, the acoustic variable on the right side of this equation is related to the pressure perturbation $p$ as $p= \rho'c_\infty^2$. However Lighthill's equation only gives the propagation of sound outside the turbulent flow and so, in principle, we cannot use it in the flow to obtain the surface pressure because $T_{ij}$ includes terms that depend on $\rho'$ \citep{Glegg2017}. Nevertheless this approach has been used to calculate surface pressure fluctuations beneath a turbulent boundary layer \citep{Grasso2019,Blake2017} and so we will review the background to this approach before introducing alternatives.

In an incompressible flow Lighthill’s equation can be used to relate the pressure perturbation to the stress tensor $T_{ij}$ by letting $c_\infty$ tend to infinity, which effectively eliminates the term that depends on the time derivative and yields the Poisson's equation
\begin{equation} 
\dfrac{\partial^2 p}{\partial x_i^2}  = -  \dfrac{\partial^2 T_{ij}}{\partial x_i \partial x_j}
\label{eq:pressTij}
\end{equation}
The right side of this equation depends on both a turbulent shear stress $\rho v_i v_j$ and viscous stress $\sigma_{ij}$ and can be expanded out as
\begin{equation} 
\dfrac{\partial^2 T_{ij}}{\partial x_i \partial x_j} = \dfrac{\partial^2 \left(\rho v_i v_j\right)}{\partial x_i \partial x_j} - \dfrac{\partial^2 \sigma_{ij}}{\partial x_i \partial x_j}
\end{equation}
However, using the definition of viscous shear stress for an incompressible flow ($\sigma_{ij} = \mu (v_{i,j}+v_{j,i})$, and $v_{i,i}=0$) we see that
\begin{equation} 
\dfrac{\partial^2 \sigma_{ij}}{\partial x_i \partial x_j} = \mu \dfrac{\partial^3 v_i}{\partial x_i \partial x_j^2} = 0
\end{equation}
so the viscous shear stress does not contribute to the source term in an incompressible flow. The only remaining quantity depends on $\rho v_i v_j$, where, because the flow is incompressible $\rho = \rho_0$ is constant. It is also possible to write this part of the source term as the product of two derivatives, so
\begin{equation} 
\dfrac{\partial^2 T_{ij}}{\partial x_i \partial x_j} = \dfrac{\partial^2 \left(\rho_0 v_i v_j\right)}{\partial x_i \partial x_j} = \rho_0 \dfrac{\partial}{\partial x_i}\left(v_i \dfrac{\partial v_j}{\partial x_j} + v_j \dfrac{\partial v_i}{\partial x_j}\right) = \rho_0 \dfrac{\partial v_j}{\partial x_i}\dfrac{\partial v_i}{\partial x_j}
\end{equation}
We can now separate the mean flow effects from the turbulent velocity fluctuations by specifying $v_i=U_i+u_i$, where $U_i$ represents the time invariant mean flow and $u_i$ the time varying part of the flow. For a two dimensional parallel mean flow in the $x_1$ direction $U_1=U(x_2)$ is only a function of $x_2$. In this case the source term becomes 
\begin{equation} 
\dfrac{\partial^2 T_{ij}}{\partial x_i \partial x_j} = 2\rho_0 \dfrac{\partial U}{\partial x_2}\dfrac{\partial u_2}{\partial x_1} + \rho_0 \dfrac{\partial u_j}{\partial x_i}\dfrac{\partial u_i}{\partial x_j}
\label{eq:TijSimplified}
\end{equation}
Note here that the factor of $2$ arises because of the repeated indices. The two parts of the source term are sometimes referred to as the mean shear-turbulence and the turbulence-turbulence interaction terms.

The important conclusion from these derivations is that the unsteady pressure in an incompressible shear flow apparently depends on two different mechanisms. The first is linear in fluctuating quantities and also depends on the mean shear. It implies that if the flow is uniform so $\partial U/ \partial x_2=0$, then there are no pressure perturbations generated by linear fluctuations, and only the non-linear terms are important. It also implies that regions of high mean shear will result in larger pressure fluctuations. The problem with this interpretation is that the mean shear and the turbulence are not necessarily uncoupled, and knowledge of both is needed in order to correctly identify the regions of the flow that cause the unsteady pressure. The non-linear term is more complicated, as it depends on the interaction of the turbulence with itself. In other words the turbulent eddies are convected by not only the mean flow but also the turbulence that randomly changes the direction of motion of the eddy. The random trajectory of the eddy results in momentum flux that must be matched by a pressure perturbation, and this is reflected in the non-linear source term. In a turbulent boundary layer both the mean shear and the turbulent shear stress are largest close to the surface, and so knowledge of both terms is needed to determine the surface pressure. Some studies suggest that the mean shear term dominates \citep{Blake2017}, but others suggest that both terms are of the same magnitude, and identifying their relative importance remains a challenge.

\section{The Surface Pressure Wavenumber Spectrum}
\label{sec:pressureSpectrum}

\subsection{The Solution to the Poisson's Equation for the Pressure}
The applications that we are concerned with require the input to be specified as the wavenumber spectrum of the surface pressure fluctuations (see \citet{Blake2017} and \citet{Grasso2019}). This is the quantity required to calculate the noise from a rough surface or a trailing edge \citep{Glegg2017}, and similarly, the vibration of a panel caused by an attached turbulent boundary layer \citep{Blake2017}. To obtain the surface pressure wavenumber spectrum we apply Fourier transforms to obtain the pressure wavenumber spectrum at a location $x_2$ as
\begin{equation}
{\dbtilde p}(x_2, k_1, k_3, \omega) = \dfrac{1}{\left(2\pi \right)^3} \int \displaylimits_{-T}^{T} \int\displaylimits_{-R}^{R} \int\displaylimits_{-R}^{R} p(\bm{x}, t) e^{i\omega t - ik_1 x_1 - ik_3 x_3} dt dx_1 dx_3
\end{equation}

Note that the pressure is now defined as a function of frequency $\omega$, the wavenumbers $k_1$ and $k_3$, and the distance from the surface $x_2$. The constants $R$ and $T$ both tend to infinity.

When this transform is applied to the Poisson's equation for the pressure perturbations in the flow (using equations~\ref{eq:pressTij} and~\ref{eq:TijSimplified}), it becomes
\begin{equation}
\dfrac{\partial^2 \dbtilde p}{\partial x_2^2} - k_s^2 \dbtilde p = \dbtilde q_{MS}(x_2, k_1, k_3, \omega) + \dbtilde q_{TT}(x_2, k_1, k_3, \omega)
\label{eq:pressurePoisson}
\end{equation}
Where $k_s=(k_1^2+k_3^2)^{1/2}$ and the source terms are defined by 
\begin{subequations}
\begin{align}
\dbtilde q_{MS}(x_2, k_1, k_3, \omega) &= -\dfrac{1}{\left(2\pi \right)^3} \int \displaylimits_{-T}^{T} \int\displaylimits_{-R}^{R} \int\displaylimits_{-R}^{R}  \left(2\rho_0\dfrac{\partial U}{\partial x_2}\dfrac{\partial u_2}{\partial x_1} \right) e^{i\omega t - ik_1 x_1 - ik_3 x_3} dt dx_1 dx_3\\
\dbtilde q_{TT}(x_2, k_1, k_3, \omega) &= -\dfrac{1}{\left(2\pi \right)^3} \int \displaylimits_{-T}^{T} \int\displaylimits_{-R}^{R} \int\displaylimits_{-R}^{R} \left(\rho_0\dfrac{\partial u_j}{\partial x_i}\dfrac{\partial u_i}{\partial x_j} \right)  e^{i\omega t - ik_1 x_1 - ik_3 x_3} dt dx_1 dx_3
\end{align}
\label{eq:sourceTerms}
\end{subequations}
The problem is now cast in the form of a second order inhomogeneous differential equation (equation~\ref{eq:pressurePoisson}) that can most conveniently be solved by using a Green's function that is the solution to
\begin{equation}
\dfrac{\partial^2 g(x_2, y)}{\partial x_2^2} - k_s^2 g(x_2,y) = \delta(x_2 - y)
\end{equation}
It is relatively easy to show that 
\begin{equation}
g(x_2, y) = -\left( \dfrac{e^{-k_s \lvert x_2 - y \rvert} + Ae^{-k_s(x_2+y)}}{2k_s} \right)
\end{equation}
where $A$ is a constant defined by the boundary conditions. The pressure is then given as
\begin{equation}
\dbtilde p(x_2) = \int\displaylimits_{0_+}^\infty \left(\dbtilde{q}_{MS}(y)  + \dbtilde{q}_{TT}(y)\right) g(x_2, y) dy
\end{equation}
The surface pressure is readily obtained by setting $y>x_2=0$, so 
\begin{equation}
\dbtilde p(0) = \int\displaylimits_{0_+}^\infty \dfrac{(1+A)e^{-k_s y}}{2k_s} \left(\dbtilde{q}_{MS}(y)  + \dbtilde{q}_{TT}(y)\right) dy
\end{equation}
To determine the constant $A$ we consider the $x_2$ component of the momentum equation on the surface, and apply the no slip boundary condition so all the flow velocities are zero on the wall so
\begin{equation}
\left[ \dfrac{\partial \dbtilde p}{\partial x_2} \right]_{x_2=0}  =  \left[ \mu \dfrac{\partial^2 \dbtilde u_2}{\partial x_2^2} \right]_{x_2=0}
\end{equation}
The surface pressure gradient therefore depends on the second derivative of the unsteady velocity normal to the wall. Without more detailed modeling this is hard to specify, but the scaling of the flow at very high Reynolds numbers would suggest that the viscous terms can be ignored, so the pressure gradient is zero at the wall. In that case the derivative of the Green's function $g(x_2,y)$ with respect to $x_2$ must be zero at $x_2=0$, when $y>0$, and so $A=1$.

\subsection{The Surface Pressure}
The analysis given in the previous section assumed that the terms on the right side of equation~\ref{eq:pressurePoisson} were source terms, which implies that they are not related to the pressure perturbation that is the dependent variable of the differential operator on the left side of the equation. This is not the case because the pressure and normal velocity are related by the momentum equation. To show this consider the source terms given by equation~\ref{eq:sourceTerms} in more detail. The mean shear part may be written as
\begin{equation}
\dbtilde q_{MS}(x_2, k_1, k_3, \omega) =  -2ik_1 \rho_0 U' \dbtilde u_2(x_2, k_1,k_3,\omega)
\label{eq:qMS}
\end{equation}
where $U'$ is the mean shear $\partial U/ \partial x_2$, and a prime will be used to denote a differentiation with respect to $x_2$. This notation will be used in the rest of the paper, but does not apply to $\rho'$ that only appears in \S\ref{sec:sourcePressure}. 

Equation~\ref{eq:qMS} shows that in the wavenumber frequency domain the mean shear turbulence interaction term is proportional to $\dbtilde u_2$, which can also be related to the pressure perturbations by the component of the momentum equation in the $x_2$ direction, given as
\begin{equation}
\rho_0\dfrac{\partial u_2}{\partial t} + \rho_0 U \dfrac{\partial u_2}{\partial x_1} + \dfrac{\partial p}{\partial x_2} - \mu \nabla^2 u_2 = -\rho_0 u_j \dfrac{\partial u_2}{\partial x_j}
\end{equation}
which, after applying the Fourier transforms defined above gives
\begin{equation}
\dbtilde p \ '  + ik_1 \rho_0 (U - c) \dbtilde u_2 - \mu \left(\dbtilde u_2'' - k_s^2 \dbtilde u_2 \right) = \dfrac{\dbtilde q_2}{ik_1}
\label{eq:pPrime}
\end{equation}
 where $c=\omega/k_1$ is the phase speed, and $q_2$ is the non-linear source term.
 
The Poisson's equation for the pressure (equation~\ref{eq:pressurePoisson}) is then re-written by substituting equation~\ref{eq:qMS} and moving the mean shear part of the source term to the left side so
\begin{equation}
\dbtilde p \ ''  - k_s^2 \dbtilde p + 2 i k_1 \rho_0 U' \dbtilde u _2 = \dbtilde q_{TT}(x_2, k_1, k_3, \omega)
\label{eq:pPrimePrime}
\end{equation}
We now have two coupled linear differential equations~\ref{eq:pPrime} and~\ref{eq:pPrimePrime} for the variables $p$ and $u_2$. In principle one cannot be solved without also solving the other. If, however, one of the variables, such as the velocity, is known then only one equation has to be solved. However, the result depends on the accuracy with which the individual variable is defined, and that is often subject to large uncertainty, and flow specific approximations. Nevertheless it is the basis for the Blake-TNO approach to obtain surface pressure \citep{Blake2017}.

\section{The Surface Pressure Calculations based on the Orr Sommerfeld Equation}

As an alternative to the formulation of the problem based on the Poisson's equation for the pressure, \citet{Landahl1967} considered the problem in terms of the inhomogeneous Orr-Sommerfeld equation. This approach treats the non-linear terms as the effective sources of all the fluctuating quantities in the boundary layer, and as such represents a complete solution to the Navier Stokes equations if the non-linear terms are known. A more detailed derivation of this approach was given by \citet{Chase1991} who started with the vorticity equation in a parallel shear flow and applied the wavenumber transforms specified in \S\ref{sec:pressureSpectrum}. Then by using Squires transformation the result was reduced to the Orr-Sommerfeld equation for two of the three components of the unsteady velocity, with the third component related to the other by the momentum equation. In this section we will re-visit Chase’s derivation of the inhomogeneous Orr-Sommerfeld equation and then show how the surface pressure fluctuations can be related to the vorticity in the flow.

\subsection{Squires Transformation}

As before, we will consider a parallel shear flow that is defined by the mean velocity $U=(U(x_2),0,0)$ in the $x_1$ direction. Superimposed on the mean flow there is a velocity perturbation, and for simplicity, we will only consider a single wavenumber frequency component, which will be specified as  
\begin{equation}
u_i(\bm{x}, t) = w_i(x_2, k_1, k_3, \omega) e^{-i\omega t + i k_1 x_1 + i k_3 x_3}
\end{equation}
The general solution is then given by the inverse Fourier transforms of each variable for each frequency and wavenumber. We are concerned with turbulent flows that are statistically stationary in time and homogeneous in the plane of the mean flow, and this requires that the frequency $\omega$ and the two wavenumbers $k_1$ and $k_3$ are real and have values in the range $\pm \infty$.

\begin{figure}
\centering
\includegraphics[width=0.4\linewidth]{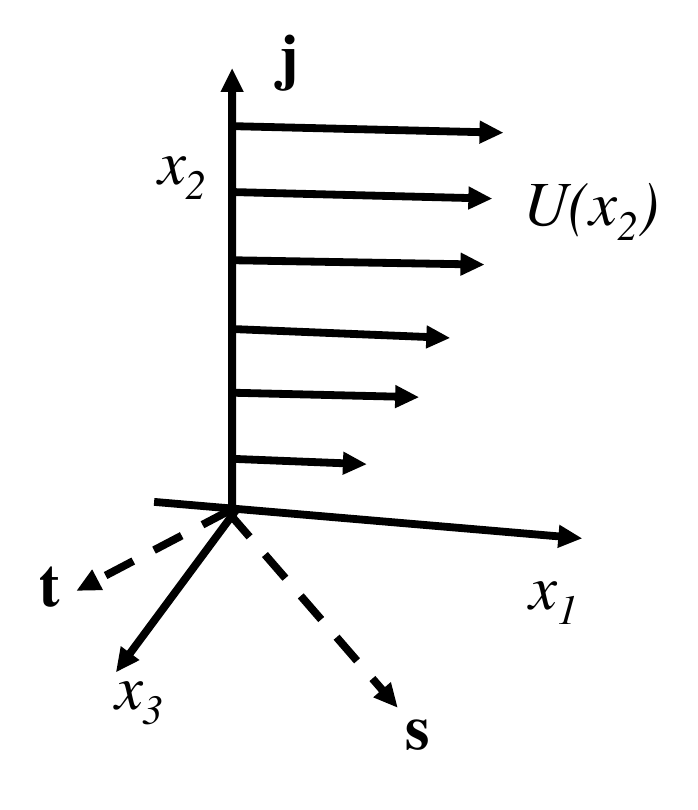}
\caption{\label{fig:wavenumberVector} The coordinates and the directions of the unit vectors $\bm s$,  $\bm j$ and $\bm t$.
}
\end{figure}

For convenience we introduce the wavenumber vector $\bm{k}=(k_1,0,k_3)$ and the unit vector in the direction of $\bm k$ which is $\bm s=(k_1/k_s,0,k_3/k_s)$ where $k_s=(k_1^2+k_3^2)^{1/2}$ (see figure~\ref{fig:wavenumberVector}). 

There are two vectors orthogonal to $\bm s$, the first is $\bm j$ that lies in the $x_2$ direction and the second is $\bm t=(-k_3/k_s,0,k_1/k_s)$ which lies in the plane of the flow. Then $k_1x_1+k_3x_3=\bm{k}\cdot\bm{x}=k_sx_s$ where $ x_s=\bm{s}\cdot\bm{x}$. 

This leads to some important simplifications for incompressible flows for which $\nabla \cdot \bm{u}=0$. It follows that
\begin{equation}
ik_1w_1 + w_2' + ik_3w_3 = ik_sw_s + w_2' = 0
\end{equation}
where $w_s$ is the component of the velocity in the direction of the unit vector $\bm s$ and we have used a prime to indicate differentiation with respect to $x_2$. It follows that the $w_s$ and $w_2$ components of the velocity are specified by the stream function $\psi (x_2)$ such that
\begin{subequations}
\begin{align}
w_s &= \psi' \\
w_2 &= -ik_s\psi
\end{align}
\end{subequations}
and the vorticity in the direction of the unit vector $\bm t$ is
\begin{equation}
\omega_t = k_s^2\psi - \psi''
\label{eq:omegat}
\end{equation} 
There is also a velocity component in the direction $\bm t$ which is defined as $w_t$ and determines the vorticity components in the directions normal to the flow and in the direction of the wave vector since
\begin{subequations}
\begin{align}
\omega_2 &= -ik_sw_t \\
\omega_s &= w_t'
\end{align}
\end{subequations}
To obtain $w_t$ we consider the momentum equation
\begin{equation}
\dfrac{\partial u_i}{\partial t} + U\dfrac{\partial u_i}{\partial x_1} + u_2 U' \delta_{i1} + \dfrac{1}{\rho_0}\dfrac{\partial p}{\partial x_i} - \nu \dfrac{\partial^2 u_i}{\partial x_j^2} = -u_j\dfrac{\partial u_i}{\partial x_j}
\end{equation}
which after applying the the Fourier transforms in time and space gives
\begin{equation}
i(k_1 U - \omega) w_i + w_2 U' \delta_{i1} + ik_ip - \nu (w_i'' - k_s^2 w_i) = -q_i
\label{eq:momentum}
\end{equation}
Evaluating this in the $\bm t$ direction gives
\begin{equation}
i(k_1 U - \omega) w_t + w_2 U't_1 - \nu (w_t'' - k_s^2 w_t) = -q_t
\end{equation}
where the pressure term has been eliminated because there is no variation of the perturbed quantities in the $\bm t$ direction since it is parallel to the wave fronts. If the non-linear terms, specified by $q_t$, are small, which implies that $U>>w_2$ and $w_s$, then this gives a second order differential equation that relates $w_t$ to $w_2$, and hence $\psi$. If the Reynolds number of the flow is large, and the viscous terms are ignored, then there is a simple algebraic relationship between $w_t$ and $w_2$, and the problem is reduced to determining a single perturbation quantity such as $w_2$ or $\psi$. 

The representation of a three dimensional flow perturbation in a parallel shear flow by a single stream function $\psi$  is known as Squires theorem, and can be found in many texts including \citet{Drazin2004}. 

\subsection{The Inhomogeneous Orr Sommerfeld Equation}
The vorticity equation can be obtained by taking the curl of the momentum equation and, if the mean vorticity is $\bm \Omega (x_2)=(0,0,\Omega)$ and the vorticity perturbation is $\bm \omega(x,t)$, then we find that,
\begin{equation}
\dfrac{\partial \bm \omega}{\partial t} + \left( \bm U \cdot \nabla \right) \bm\omega -  \left(\bm \omega \cdot \nabla \right) \bm U + \left(\bm u \cdot \nabla \right) \bm \Omega - \left(\bm \Omega \cdot \nabla \right) \bm u - \nu \nabla^2 \bm \omega = \left(\bm\omega \cdot \nabla \right) \bm u - \left( \bm u \cdot \nabla \right) \bm \omega
\end{equation}
For the case of the parallel shear flow with a harmonic perturbation, as discussed above, the vorticity in the $\bm t=(-k_3/k_s,0,k_1/k_s)$ direction is given by $\bm \omega \cdot \bm t = \omega_t(x_2)\exp(-i\omega t+i\bm k \cdot \bm x)$ and is the solution to the equation
\begin{equation}
ik_1\left(U - c\right) \omega_t + \omega_2 U' k_3/k_s + w_2 \Omega' k_1/k_s - i k_3 \Omega w_t - \nu \left(\omega_t'' - k_s^2 \omega_t \right) = \theta_t
\end{equation}
where the non linear terms are specified by $\theta_t$ (see \citet{Chase1991}). Since the mean vorticity is related to the mean flow velocity by $\Omega = -U'$ and, as was shown above, $\omega_2 = -ik_s w_t$, the second and fourth terms cancel so we obtain
\begin{equation}
ik_1\left(U - c\right) \omega_t - w_2 U'' k_1/k_s - \nu \left(\omega_t'' - k_s^2 \omega_t \right) = \theta_t
\label{eq:omegatDiffeq}
\end{equation}
Noting that $w_2=-ik_s\psi$, equation~\ref{eq:omegatDiffeq} and equation~\ref{eq:omegat} give two coupled differential equations for the vorticity and the stream function,
\begin{subequations}
\begin{align}
ik_1(U-c)\omega_t + i k_1 U'' \psi - \nu (\omega_t'' - k_s^2\omega_t) &= \theta_t \\
\psi'' - k_s^2 \psi &= -\omega_t
\end{align}
\end{subequations}
and when these are combined we obtain 
\begin{equation}
-ik_1\left(U - c\right) (\psi'' - k_s^2\psi) + i k_1 \psi U'' + \nu (\psi'''' - 2k_s^2\psi'' + k_s^4\psi) = \theta_t
\label{eq:omegatDiffeqSecond}
\end{equation}
This is the inhomogeneous form of the Orr-Sommerfeld (OS) equation, with the non-linear terms included on the right side as source terms. It provides the complete solution for the flow given the relationships in the previous section. The non-linear term is, strictly, part of the solution. However, it is a function of perturbations at different wavenumbers, and these are uncorrelated with the perturbations at the frequency and wavenumbers being considered. We therefore argue that the non-linear terms are effectively source terms for the perturbations since they do not interfere with the wave motion that is described by the terms on the left side of the equation.

We normalize all the variables in these equations based on the outer lengthscale $\delta$ and the outer flow speed $U_0$ so they take the form
\begin{subequations}
\begin{align}
(U-c)\omega_t + U'' \psi + i \epsilon (\omega_t'' - \alpha^2\omega_t) &= \theta_t \\
\psi'' - \alpha^2 \psi &= -\omega_t
\end{align}
\label{eq:OSnormalized}
\end{subequations}
Where $\epsilon = \nu/k_1 U_o \delta^2 = 1/\alpha_1 Re$,  $\alpha_1=k_1 \delta$ and $\alpha=k_s \delta$. Since $Re$ is the outer flow Reynolds number we can always assume that this is a large parameter and $\epsilon<<1$. All functions are dependent on $y=x_2/\delta$  and a prime donates a differentiation with respect to non-dimensional distance from the wall. 

\subsection{The Surface Pressure}
The surface pressure $p_s$ is obtained from the momentum equation~\ref{eq:momentum} evaluated at the wall in the $\bm s$ direction. Because of the no slip and non penetration boundary condition this simplifies to
\begin{equation}
i k_s p_s = \nu \left[(w_s'' - k_s^2w_s)\right]_{x_2 = 0} = -\nu \omega_t'(0)
\end{equation}
Non-dimensionalizing this based on outer units, with pressure normalized on $\rho U_0^2,$ gives the wall pressure as
\begin{equation}
\left(\dfrac{\rho_0 U_0^2}{\delta}\right)i\alpha p_s = -\omega_t'(0) \left(\dfrac{\mu U_0}{\delta^2} \right)
\end{equation}
So
\begin{equation}
p_s = i\epsilon \omega_t'(0) = i \omega_t'(0)/\alpha_1 Re
\label{eq:ps}
\end{equation}
This gives a relatively simple set of equations~\ref{eq:OSnormalized},  and~\ref{eq:ps} that can be solved for the surface pressure providing that the non-linear terms are known. These are specified by \citet{Chase1991} as
\begin{equation}
\theta_t = \left(\dbtilde T_{j2} '' + \alpha^2 \dbtilde T_{j2}\right) s_j  + i \alpha \left(\dbtilde T_{ij} ' s_i s_j - \dbtilde T_{22} ') \right)
\end{equation}
where $\dbtilde T_{ij}$ is the frequency space transform of $u_i u_j/U_o^2$.

There are several different methods for solving the inhomogeneous OS equation. \citet{Landahl1967} used a modal approach solving for the eigenvalues of the complex phase speed $c$ and the associated orthogonal modes that match the boundary conditions at the surface and in the outer flow. However it was noted  that the numerical errors associated with this approach at high Reynolds number could become large. \citet{Chase1991} considered the asymptotic case that the wavenumber $k_1$ tended to zero. He used the method of variation of parameters to solve the inhomogeneous OS equation and was thus able to specify a continuous solution for all wavenumbers and frequencies. However if either of these approaches are taken the non-linear terms and the upstream boundary conditions need to be specified to obtain a complete solution, and as was noted above, these are difficult to both measure and model. For that reason we will reconsider the problem by making use of the second of equation~\ref{eq:OSnormalized} which relates the vorticity component $\omega_t$ to the velocity field. If this equation can be solved in terms of the vorticity within the flow then we can use the momentum equation~\ref{eq:ps} to obtain the surface pressure. However the vorticity next to the surface is dominated by the viscous shear stress which is needed to ensure the no slip boundary condition is met. This generates vorticity at the boundary, which must extend into the flow since it forms the viscous sublayer. In addition to this the turbulence outside the viscous sublayer generates vorticity of it’s own, which, by the Biot Savort law must induce a flow parallel to the surface. It is the conjecture of this paper that the flow induced by the turbulence outside the viscous sublayer, the `outer' vorticity, is exactly cancelled by the viscous shear stress at the wall, whose gradient in the wall normal direction is exactly equal to the pressure gradient at the wall. If this is the case then the surface pressure can be defined simply in terms of the outer vorticity and this simplifies the modeling and measurement of the flow characteristics that create surface pressure fluctuations.

\subsection{Separating the Inner and Outer Vorticity Terms}

The next step in the analysis is to split the vorticity in the boundary layer into a part that defines the surface pressure and a part that allows for the wall shear stress. On this basis we specify
\begin{subequations}
\begin{align}
\omega_t &= \omega_p + A \omega_w \\
\psi &= \psi_p + A \psi_w
\end{align}
\end{subequations}
Where $A$ is a constant. In this equation $\omega_p$ is the solution to the inhomogeneous OS equation that matches the boundary conditions that the vorticity tends to zero far from the wall but does not impact the wall shear stress. Hence the vorticity is zero at the wall, so $\omega_p(0)=0$. The vorticity $\omega_w$ is the vorticity in the viscous sublayer that ensures the no slip boundary condition, and, since the non-linear terms are zero at the wall, it will be a solution to the homogeneous OS equation that decays at infinity and has the boundary condition at the wall of $\omega_w(0)=1$. The corresponding streamfunctions are solutions to the Poisson's equation given by the second of equation~\ref{eq:OSnormalized} and so can have their own boundary conditions. These are chosen so that the stream function tends to zero at large distances from the wall, and matches the non penetration boundary condition $\psi(0)=\psi_p(0)=\psi_w(0)=0$. It follows then that the solutions to equation~\ref{eq:OSnormalized} will be of the form
\begin{subequations}
\begin{align}
\psi_p(y) &= \dfrac{1}{2\alpha} \int\displaylimits_{0_+}^\infty \omega_p(y_0) \left\{ e^{-\alpha \lvert y - y_0 \rvert} - e^{-\alpha (y+y_0)}\right\} d y_0 \\
\psi_w(y) &= \dfrac{1}{2\alpha} \int\displaylimits_{0}^\infty \omega_w(y_0) \left\{ e^{-\alpha \lvert y - y_0 \rvert} - e^{-\alpha (y+y_0)}\right\} d y_0
\end{align}
\end{subequations}

To match the no slip condition at the wall the constant $A$ is required to be
\begin{equation}
A = -\psi_p'(0)/\psi_w'(0)
\end{equation}
and so
\begin{equation}
A = \dfrac{1}{\psi_w'(0)} \int\displaylimits_{0_+}^\infty \omega_p(y_0) e^{-\alpha y_0} dy_0
\end{equation}
which represents the integral of the outer vorticity in the region $y_0>0$ with an exponential weighting. 

We can then obtain the pressure spectrum from equation~\ref{eq:ps} as
\begin{equation}
p_s = i\epsilon (A \omega_w'(0) + \omega_p'(0) ) = \dfrac{i \epsilon \omega_w'(0)}{\psi_w'(0)} \int\displaylimits_{0_+}^\infty \omega_p(y_0) e^{-\alpha y_0} dy_0 + i\epsilon \omega_p'(0))
\label{eq:pressureSpectrum}
\end{equation}
In the viscous sublayer the mean velocity gradient is constant, and so $U''=0$. Consequently the first equation given by equation~\ref{eq:OSnormalized} is uncoupled from the second. Furthermore, the mean flow speed tends to zero so we can assume, at least in the first instance when $c>>U(y)$, that $\omega_w\sim\exp(-\alpha_0 y)$ where
\begin{equation}
\alpha_0^2 = \alpha^2 - ic/\epsilon
\end{equation}
Note that, for the non-dimensional frequency $\omega$,  $c=\omega/\alpha_1$ so $c/\epsilon=(\omega/\alpha_1)\alpha_1 Re = \omega Re$ and we see that we also have the relationship 
\begin{equation}
\alpha_0^2 = \alpha^2 - i\omega Re
\end{equation}
Since $\omega$ is of order one in typical boundary layer flows the near wall decay rate of the shear stress vorticity is very rapid at high Reynolds number. 

The velocity perturbations induced by the shear stress at the wall can then be assumed to be
\begin{subequations}
\begin{align}
\psi_w(y) &= \dfrac{1}{(\alpha_0^2 - \alpha^2)} \left\{ e^{-\alpha_0  y} - e^{-\alpha y}\right\} \\
\psi_w'(y) &= \dfrac{-1}{(\alpha_0^2 - \alpha^2)} \left\{\alpha_0 e^{-\alpha_0  y} - \alpha e^{-\alpha y}\right\}
\end{align}
\end{subequations}
and
\begin{equation}
\omega_w(y) = e^{-\alpha_0 y}
\end{equation}
Using these solutions gives the constants needed in equation~\ref{eq:pressureSpectrum} as
\begin{equation}
\dfrac{i \epsilon \omega_w'(0)}{\psi_w'(0)} = \dfrac{-i \epsilon (\alpha_0^2 - \alpha^2)\alpha_0}{\alpha_0 - \alpha} = \dfrac{-i \epsilon \left(\dfrac{-ic}{\epsilon}\right)\alpha_0}{\alpha_0 - \alpha} 
\end{equation}
Hence the surface pressure is
\begin{equation}
p_s = -\dfrac{c \alpha_0}{\alpha_0 - \alpha} \int\displaylimits_{0_+}^\infty \omega_p(y_0) e^{-\alpha y_0} dy_0 + i\epsilon \omega_p'(0)
\label{eq:surfacePressure}
\end{equation}
For turbulent flows at high Reynolds number where $\alpha_0 >> \alpha$ and $\epsilon<<1$ this simplifies to
\begin{equation}
p_s = -c  \int\displaylimits_{0_+}^\infty \omega_p(y_0) e^{-\alpha y_0} dy_0
\label{eq:surfacePressurehiRe}
\end{equation}
This is the key result of this paper because it shows that the surface pressure wavenumber spectrum can be defined by the integral of the outer vorticity as a function of distance from the wall, with an exponential weighting factor. In obtaining this result the solution to the OS equation next to the wall assumed that the mean flow speed was much less than the phase speed, but this assumption is not necessary and more accurate solutions can be obtained if needed using either Airy functions or the WKB method (see \citet{Drazin2004}) for a linear near wall mean velocity profile. We still, however, need to specify the outer vorticity and so in the next section we will use a DNS calculation of a channel flow to first verify the results given by equation~\ref{eq:surfacePressurehiRe}, and then guide the modeling of the outer vorticity. Following that we will develop suitable models for the vorticity wavenumber spectrum that can be used to obtain the surface pressure spectrum in terms of the flow variables.

\section{Verification of Results using DNS}
In order to obtain the surface pressure wavenumber spectrum using equation~\ref{eq:surfacePressure} it is necessary to split the vorticity into the part defined by the wall shear stress and the part that is caused by the non-linear source terms and upstream boundary conditions. The vorticity associated with the wall shear stress has been modeled as $A \omega_w(y)= A \exp(-\alpha_0 y)$ and we have specified that $\omega_p(0)=0$. It follows then that $A$ is equal to the vorticity at the wall. By using a DNS calculation we can there verify that the modeling based on equation~\ref{eq:surfacePressure} and~\ref{eq:surfacePressurehiRe} is correct by evaluating 
\begin{equation}
\omega_p(y) = \omega_t(y) - \omega_t(0) e^{-\alpha_0 y}
\label{eq:omegaPy}
\end{equation}
from the DNS calculation and then evaluating equation~\ref{eq:surfacePressurehiRe} to verify that the pressure spectrum obtained from the DNS is reproduced. Given that, we can then consider the properties of $\omega_p(y)$ to gain insights into the mechanisms of surface pressure fluctuations in the flow.

\subsection{The DNS Calculations}

The turbulent flow data used in this work was generated using Direct Numerical Simulation (DNS) of a periodic channel flow. The incompressible Navier-Stokes equations were solved on a staggered grid using a second order finite difference scheme and the second order semi-implicit Crank-Nicolson scheme~\citep{Desjardins2008}. The flow was driven in the channel by imposing a constant pressure gradient in the streamwise direction. The friction Reynolds number was set to be $Re_\tau = u_{\tau} (L_y/2)/\nu \approx 300$, where $u_\tau = \sqrt{\tau/\rho}$ is the friction velocity and $\tau = \mu U'$ is the surface shear stress. Periodic boundary conditions were used in the streamwise and spanwise directions, and the no-slip boundary condition was enforced at the top and bottom walls. A stretched Cartesian grid was used in the wall-normal direction to resolve the viscous length scales close to the wall. The minimum grid cell height $\Delta y$ was $0.03\delta^+$ next to the wall, and the maximum $\Delta y$ was $2.4\delta^+$ in the core region, with the cell height stretched using a hyperbolic tangent function. Here, $\delta^+ = \nu/u_{\tau}$ is the viscous length scale, $\nu = \mu/\rho$ is the kinematic viscosity, $\mu$ is the dynamic viscosity, and $\rho$ is the fluid density. The grid cell sizes were kept uniform in the streamwise and spanwise directions ($\Delta x=\Delta z = 3.5\delta^+$). Upon reaching a statistically stationary state, the vorticity components were calculated and snapshots separated by short time intervals of $3e$-$5 t^+$ were saved (where $t^+=\delta^+/u_\tau$ is the viscous time scale) in order to compute time-derivatives. Additionally the wall-pressure values were stored for computing the relevant spectra in \S~\ref{subsec:pressVortDNS}.

\subsection{Surface Pressure and Vorticity from DNS}
\label{subsec:pressVortDNS}

The DNS calculation provides the pressure, velocity and vorticity in the flow as a function of space and time. In contrast, the theory given above is in the wavenumber frequency domain and thus the DNS data must be transformed into Fourier space as a function of $k_1$, $k_3$ and $\omega$ to give a direct comparison with equation~\ref{eq:surfacePressurehiRe}. The size of the data set for a well resolved time-series is expected to be on the order of 10 terabytes, so to reduce the computational effort we will only consider the surface pressure wavenumber spectrum in space at one time instance. To obtain this, we need to take the Fourier transform of the data with respect to $k_1$ and $k_3$, and consider the inverse transform of equation~\ref{eq:surfacePressurehiRe} as a function of frequency, so at time $t=0$ the surface pressure wavenumber spectrum $p_s^{(t)} (\alpha_1, \alpha_3,t)$  is
\begin{equation}
p_s^{(t)} (\alpha_1, \alpha_3,t) = -\int\displaylimits_{-\infty}^\infty c \int\displaylimits_{0_+}^\infty \omega_p(y_0, \alpha_1, \alpha_3, \omega) e^{-\alpha y_0 - i\omega t} dy_0 d\omega
\end{equation}
Without knowing the frequency spectrum of the vorticity this expression cannot be evaluated exactly, but in the high Reynolds number limit we can use the relationship that $c=\omega/\alpha_1$ and take the limit that $\alpha_0>>\alpha$ to give
\begin{equation}
p_s^{(t)} (\alpha_1, \alpha_3,t) = -\int\displaylimits_{-\infty}^\infty \dfrac{i\omega}{i \alpha_1} \int\displaylimits_{0_+}^\infty \omega_p(y_0, \alpha_1, \alpha_3, \omega) e^{-\alpha y_0 - i\omega t} dy_0 d\omega
\end{equation}
The remaining term in the inverse transform over frequency is then the outer vorticity, which can be evaluated to give
\begin{equation}
p_s^{(t)} (\alpha_1, \alpha_3,t) = \dfrac{1}{i\alpha_1} \dfrac{\partial}{\partial t} \int\displaylimits_{0_+}^\infty \omega_p(y_0, \alpha_1, \alpha_3, t) e^{-\alpha y_0} dy_0
\label{eq:psTimeDeriv}
\end{equation}
However, to obtain the outer vorticity as a function of time we must correct the actual vorticity using equation~\ref{eq:omegaPy}, and this includes a factor of $\alpha_0$ which is frequency dependent. The inner vorticity, given by the second term in equation~\ref{eq:omegaPy} is therefore weighted by a factor $\exp(-(\alpha^2 + i\omega Re)^{1/2}y)$. This function is sharply peaked at zero frequency, but because the fluctuating vorticity on the surface $\omega_t(0, \alpha_1, \alpha_3, \omega)$ must be zero at zero frequency the peak is eliminated. The inverse Fourier transform of $\omega_t(0, \alpha_1, \alpha_3, \omega) \exp(-(\alpha^2 + i\omega Re)^{1/2}y)$ is then approximated by
\begin{equation}
\omega_t(0, \alpha_1, \alpha_3,t) \exp(-(\alpha^2 + i\omega_m Re)^{1/2}y)
\end{equation}
where $\omega_m$ is the peak frequency in the vorticity spectrum. This approximation is most accurate at the wall, where $y=0$ but tends to amplify high frequency fluctuations in the inner vorticity when $y>0$. It will be most accurate near the spectral peak of the fluctuations where most energy is contained. The outer vorticity in equation~\ref{eq:psTimeDeriv} is empirically dependent on the choice of $\omega_m$ and the sensitivity to this parameter choice will be discussed below.  

\begin{figure}
\centering
\includegraphics[width=0.8\linewidth]{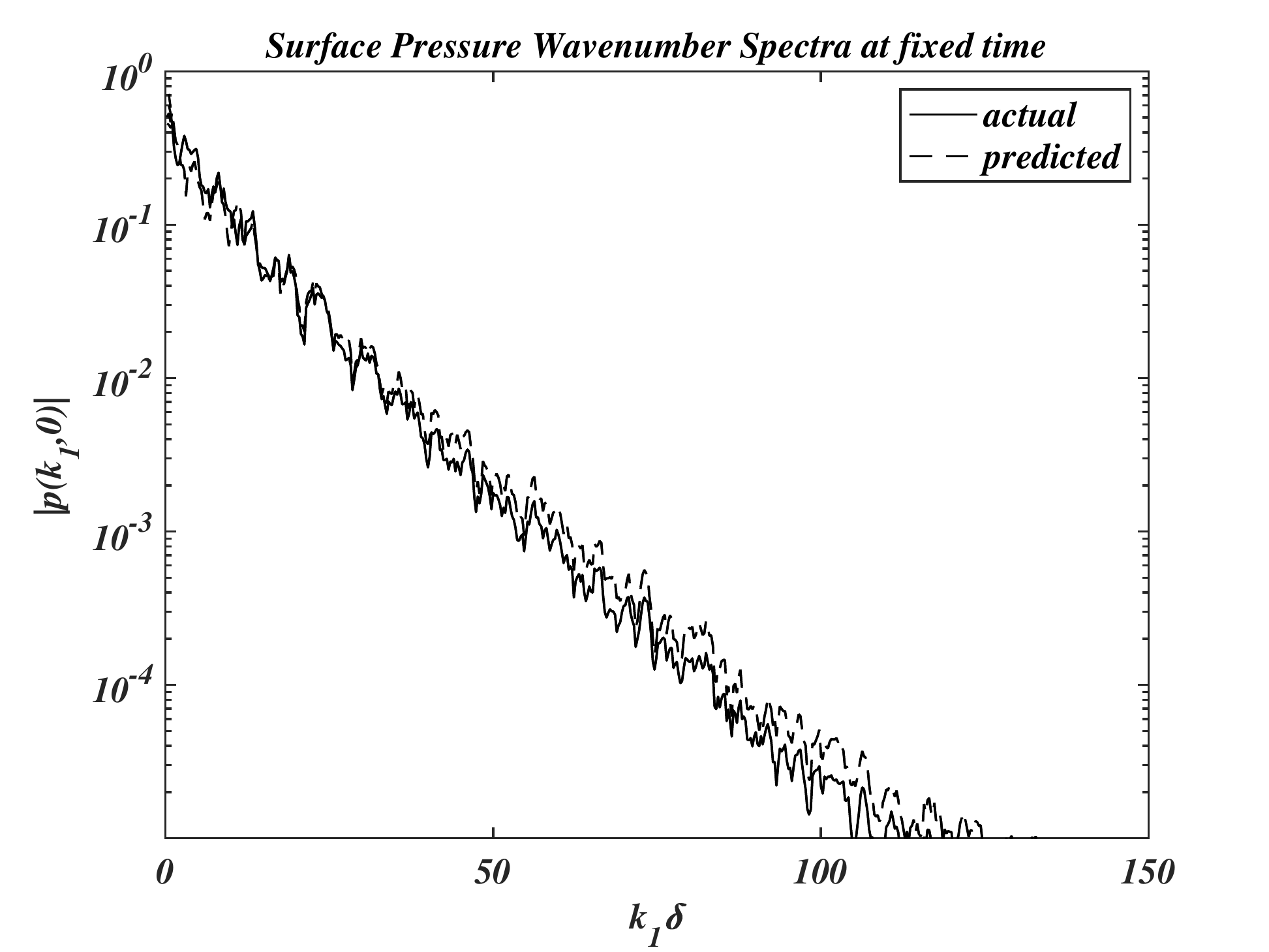}
\caption{\label{fig:pressureSpectrum} The prediction of surface pressure spectra from DNS data using a non-dimensional frequency of $\omega_m=3$.}
\end{figure}

Figure~\ref{fig:pressureSpectrum} shows the comparison between the surface pressure wavenumber spectrum at a fixed time with the surface pressure spectrum predicted from equation~\ref{eq:psTimeDeriv} assuming that $\omega_m=3$ The result is shown for a wavenumber of $k_3=0$ so $\omega_t=\omega_3$ and is obtained by integrating the $\omega_3$ component of the vorticity over the spanwise direction $x_3$. The agreement is seen to be excellent using this value of $\omega_m$. Increasing $\omega_m$ to $4$ gives a level that is on the order $1 dB$ lower than that shown at low wavenumbers, but a better fit at high wavenumbers. Decreasing $\omega_m$ to $2$ gives a level that is on the order $2 dB$ higher than that shown, with a larger error at high wavenumbers. These variations indicate that the results are not very sensitive to the choice of $\omega_m$.

\begin{figure}
\centering
\includegraphics[width=0.8\linewidth]{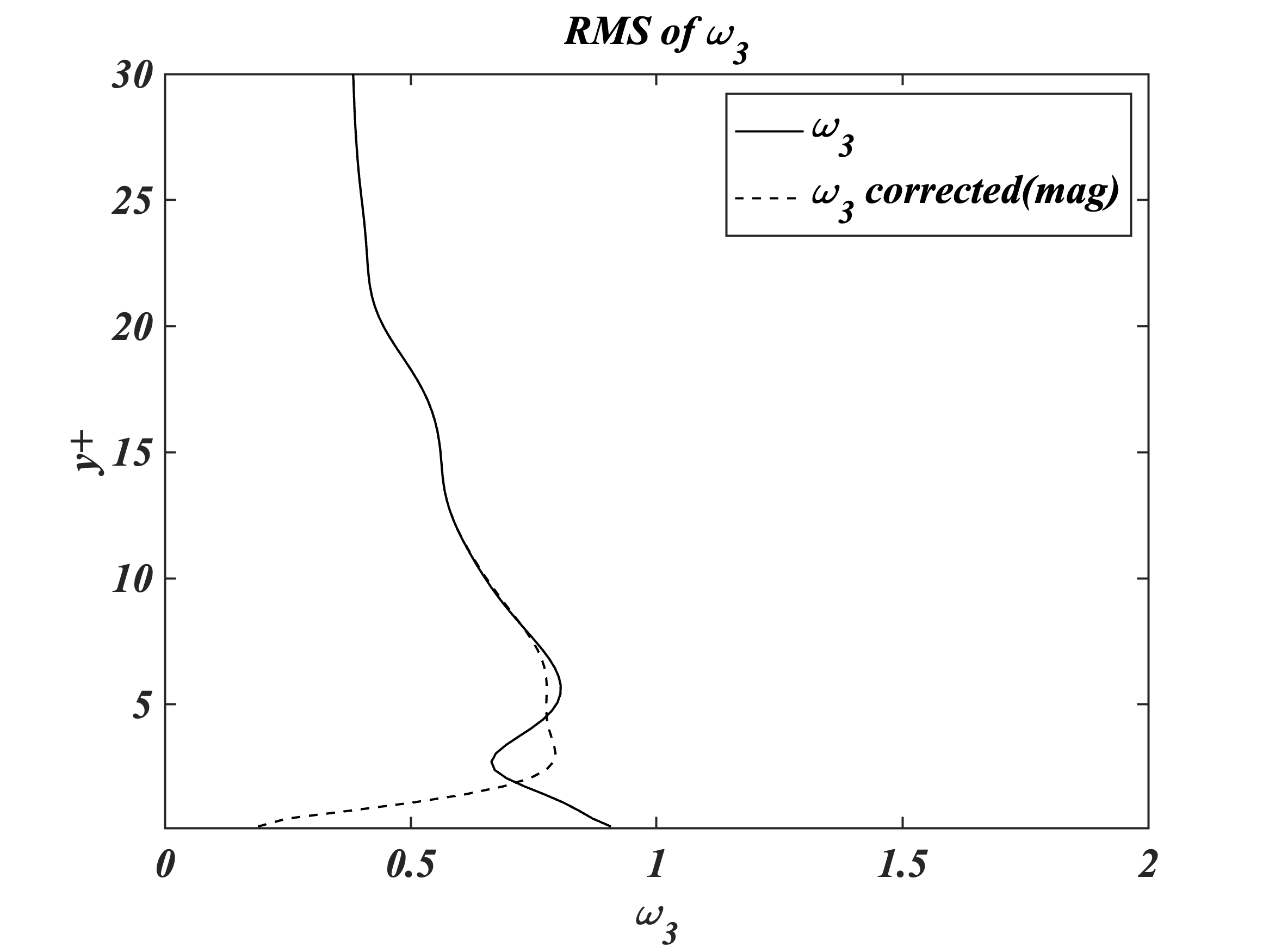}
\caption{\label{fig:vorticity} The rms of the spanwise vorticity as a function of distance from the wall.}
\end{figure}

We can also evaluate the distribution of the rms vorticity using equation~\ref{eq:omegaPy}, which is shown in Figure~\ref{fig:vorticity}. We conclude that the shear stress correction is limited to less than 10 wall units from the wall. The reason that the corrected rms value is not exactly zero at the wall is because only mesh points inside the domain have been used.

\section{The Surface Pressure Spectrum}
The theory given in the previous sections relates the instantaneous surface pressure wavenumber frequency spectra to the vorticity in the flow. In most applications we need the spectral averaged wavenumber frequency spectrum and in this section we will consider a model for this quantity based on the wavenumber frequency spectrum of the vorticity in the flow. We will only consider high Reynolds number turbulent boundary layers as distinct from the DNS of the channel flow discussed above. As a consequence we will work with dimensional units and scales that are several order of magnitude greater than the DNS calculations.

For a time stationary flow that is homogeneous in the $x_1$ and $x_3$ directions we can define the wavenumber surface pressure spectrum $\Phi_{pp}(\omega,k_1,k_3)$ and the pressure spectrum $S_{pp}(\omega)$ at a point using
\begin{subequations}
\begin{align}
{\Phi}_{pp}(\omega, k_1, k_3) &= \dfrac{\pi^3}{R^2 T} E\left[\lvert \tilde{p}(\omega, k_1, k_3) \rvert^2 \right] \\
\hat{\Phi}_{pp}(\omega, k_1) &= \int\displaylimits_{-\infty}^\infty {\Phi}_{pp}(\omega, k_1, k_3) dk_3 \\
S_{pp}(\omega) &= \int\displaylimits_{-\infty}^\infty \hat{\Phi}_{pp}(\omega, k_1) dk_1
\end{align}
\end{subequations}
Then by using equation~\ref{eq:surfacePressurehiRe} we obtain
\begin{equation}
\Phi_{pp}(\omega, k_1, k_3) = \left\lvert {\rho_0 c} \right \rvert^2  \int\displaylimits_{0}^\infty\int\displaylimits_{-x_2}^\infty e^{-k_s(2x_2+z_2)} \Omega_{tt}(\omega, k_1, k_3, x_2, z_2) dx_2 dz_2
\label{eq:phiPP}
\end{equation}
where
\begin{equation}
\Omega_{tt}(\omega, k_1, k_3, x_2, z_2) = \dfrac{\pi^3}{R^2 T} Ex\left[ \dbtilde \omega_t(\omega, k_1, k_3, x_2) \dbtilde \omega_t^*(\omega, k_1, k_3, x_2+z_2)\right]
\label{eq:omega33}
\end{equation}
We can also relate the vorticity spectrum to its correlation function. If the vorticity correlation lengthscale in the vertical direction is small then we can assume that each correlated vortex element is convected with the local mean flow speed and model the wavenumber spectrum as
\begin{align}
	\Omega_{ij}(\omega, k_1, k_3, x_2, z_2) &= \dfrac{1}{(2\pi)^3}\int\displaylimits_{-\infty}^{\infty}\int\displaylimits_{-\infty}^{\infty}\int\displaylimits_{-\infty}^{\infty}W_{ij}(x_2, z_1 - U(x_2)\tau, z_2, z_3)e^{i\omega \tau - i k_1 z_1 - i k_3 z_3} d\tau dz_1 dz_3 \\
	&= \dfrac{\delta(\omega - k_1U(x_2))}{(2 \pi)^2}\int\displaylimits_{-\infty}^{\infty}\int\displaylimits_{-\infty}^{\infty}W_{ij}(x_2, z_c, z_2, z_3) e^{-ik_1 z_c - i k_3 z_3} dz_c dz_3
\end{align}
where $W_{ij}$ is the vorticity correlation function in the moving frame. It is shown in \citet{Glegg2017} that measurements of boundary layer turbulence velocity fluctuations are well modelled by homogeneous isotropic turbulence models, such as the Von Karman or Liepmann spectrum, provided the correct length scales are chosen. Furthermore, \citet{Batchelor1999} shows that in homogeneous isotropic flows the vorticity spectrum is related to the velocity wavenumber spectrum by $\Omega_{ij}(k_1, k_2, k_3) = k^2 \Phi_{ij}(k_1, k_2, k_3)$. On this basis we could assume that
\begin{equation}
	W_{ij}(x_2, z_c, z_2, z_3) = \int\displaylimits_{-\infty}^{\infty}\int\displaylimits_{-\infty}^{\infty}\int\displaylimits_{-\infty}^{\infty}\Phi_{ij}(x_2, k_1, k_2, k_3) k^2 e^{i k_1 z_c + ik_2 z_2 + i k_3 z_3} dk_1 dk_2 dk_3 \\
\end{equation}
and specify the turbulence wavenumber spectrum in terms of the energy spectrum, so
\begin{equation}
	\Phi_{ij}(x_2, k_1, k_2, k_3) = \left(\delta_{ij} - \dfrac{k_i k_j}{k^2}\right) \dfrac{E(x_2, k)}{4\pi k^2}
\end{equation}
where $k^2=k_1^2+k_2^2+k_3^2$. However, this approach leads to non-convergent integrals because the energy spectrum asymptotes to $1/k^{5/3}$ for a Von Karman turbulence model, and $1/k^2$ for a Liepmann model. For the integrals over wavenumber to converge the energy spectrum must decay as $1/k^3$ or greater. To address this issue we will define the vorticity energy spectrum that has the characteristic
\begin{equation}
	W_{ij}(x_2, z_c, z_2, z_3) = \int\displaylimits_{-\infty}^{\infty}\int\displaylimits_{-\infty}^{\infty}\int\displaylimits_{-\infty}^{\infty}\left(\delta_{ij} - \dfrac{k_i k_j}{k^2}\right) \dfrac{E_v(x_2, k)}{4\pi} e^{i k_1 z_c + ik_2 z_2 + i k_3 z_3} dk_1 dk_2 dk_3 \\
\end{equation}

where the mean square vorticity $\omega_{rms}^2(x_2)$ is given by the integral of the vorticity energy spectrum $E_v(x_2,k)$ as
\begin{equation}
	\omega_{rms}^2(x_2) = 2  \int\displaylimits_{0}^\infty  E_v(x_2, k) k^2 dk
\end{equation}
and so

\begin{equation}
	\Omega_{tt}(\omega, k_1, k_3, x_2, z_2) = \dfrac{\delta(\omega - k_1U(x_2))}{4 \pi}\int\displaylimits_{-\infty}^{\infty} E_v(x_2, k) e^{ik_2z_2} dk_2
	\label{eq:OmetaTTfinal}
\end{equation}
and the wavenumber spectrum of the pressure is, since $c=\omega/k_1$
\begin{equation}
	\Phi_{pp}(\omega, k_1, k_3) = \dfrac{\rho^2}{4\pi} \int\displaylimits_{0}^\infty \int\displaylimits_{-x_2}^\infty \int\displaylimits_{-\infty}^\infty e^{-2k_s x_2} U^2(x_2) E_v(x_2, k) \delta(\omega-k_1 U(x_2)) e^{ik_2z_2-k_sz_2}dx_2 dz_2 dk_2
\end{equation}
The final evaluation of this integral depends on the modeling of the vorticity energy spectrum,which we will assume has a simple exponential decay determined by the viscous scale $L_v$ which is to be determined. On this basis we define
\begin{equation}
	E_v(x_2, k)=L_v^3 \omega_{rms}^2(x_2) e^{-kL_v} 
\end{equation}
which corresponds to a turbulent energy spectrum of $E(k)=E_v(k)/k^2$ with an additional exponential decay to account for the small scales. However these relationships are only orders of magnitude and it does not follow that the turbulent kinetic energy of the turbulence is proportional to the mean square vorticity as a function of height above the wall.

The integral over $k_2$ in equation~\ref{eq:OmetaTTfinal} is now given as
\begin{equation}
L_v^3\omega_{rms}^2(x_2) \int\displaylimits_{-\infty}^\infty    e^{ik_2z_2-kL_v} dk_2
\end{equation}
and in the limit that $L_v$ tends to zero this becomes
\begin{equation}
2\pi L_v^3\omega_{rms}^2(x_2) \delta(z_2)
\end{equation}
and it follows that
\begin{equation}
	\Phi_{pp}(\omega, k_1, k_3) = \dfrac{\rho^2}{2} \int\displaylimits_{0}^\infty  e^{-2k_s x_2} U^2(x_2) L_v^3\omega_{rms}^2(x_2) \delta(\omega-k_1 U(x_2)) dx_2
	\label{eq:phiPP2}
\end{equation}

The wavenumber frequency spectrum at a point is then

\begin{equation}
	\Phi_{pp}(\omega, k_1) = \dfrac{\rho_0^2 }{2}  \int\displaylimits_{-\infty}^\infty \left( e^{-2k_s y_c} L_v^3 \omega_{rms}^2(y_c) U^2(y_c)/k_1 U'(y_c)\right) dk_3
	\label{eq:phiPP3}
\end{equation}
Where $y_c$ is the height where $\omega=k_1 U(y_c)$ . The spectrum is given by
\begin{equation}
	S_{pp}(\omega) = \rho^2 \int \displaylimits_{0}^\infty  \left[ \int\displaylimits_{-\infty}^\infty e^{-2k_s x_2}  L_v^3 \omega_{rms}^2(x_2) U(x_2) dk_3 \right]_{k_1 = \omega/U(x_2)} dx_2
	\label{eq:SppNew}
\end{equation}

and the non-dimensional one sided spectrum
\begin{equation}
	\dfrac{G_{pp}(\omega)U_o}{(\rho_ou_\tau^2)^2\delta} = \left(\dfrac{U_o}{u_\tau}\right)^4 \int \displaylimits_{0}^\infty  \left[ \int\displaylimits_{-\infty}^\infty e^{-2k_s x_2}  \left(\dfrac{L_v}{\delta}\right)^3 \left(\dfrac{\omega_{rms}(x_2) \delta}{U_o} \right)^2 \dfrac{U(x_2)}{U_o} dk_3 \right]_{k_1 = \omega/U(x_2)} dx_2
	\label{eq:Gpp}
\end{equation}

In order to implement this model we need to specify the distribution of the mean square vorticity in the boundary, the mean velocity profile and the viscous length scale $L_v$. The mean velocity profile and the mean square vorticity profile is readily available from the DNS calculations, but the viscous length scale has to be estimated. Figure 4 shows the actual distribution of rms vorticity and a model that captures the near wall behaviors defined as
\begin{equation}
\omega_{rms}^2 (x_2 )=\left(\dfrac{\pi U_o^2}{100\delta^2}\right) x_2^+ exp(-x_2^+/10)
\label{eq:omegaRMSsqrd}
\end{equation}

The model correctly specifies the near wall vorticity, but underestimates the vorticity in the outer layer. 

Figure 5 shows the spectrum calculated using this profile and the rms vorticity profile from the DNS. In both cases the actual mean velocity profile was used, and the length scale $L_v=11$ wall units. The results are compared to the Goody model which is a compilation of experimental measurements made at much higher Reynolds number and extrapolated to the Reynolds number of the DNS. Clearly the spectral shape is well matched apart from at high frequencies where the Goody model tends to role off faster than the prediction. The level was found to be very sensitive to the choice of $L_v$, but, as will be discussed below the choice of eleven wall units is not unreasonable.

\begin{figure}
\centering
\includegraphics[width=0.8\linewidth]{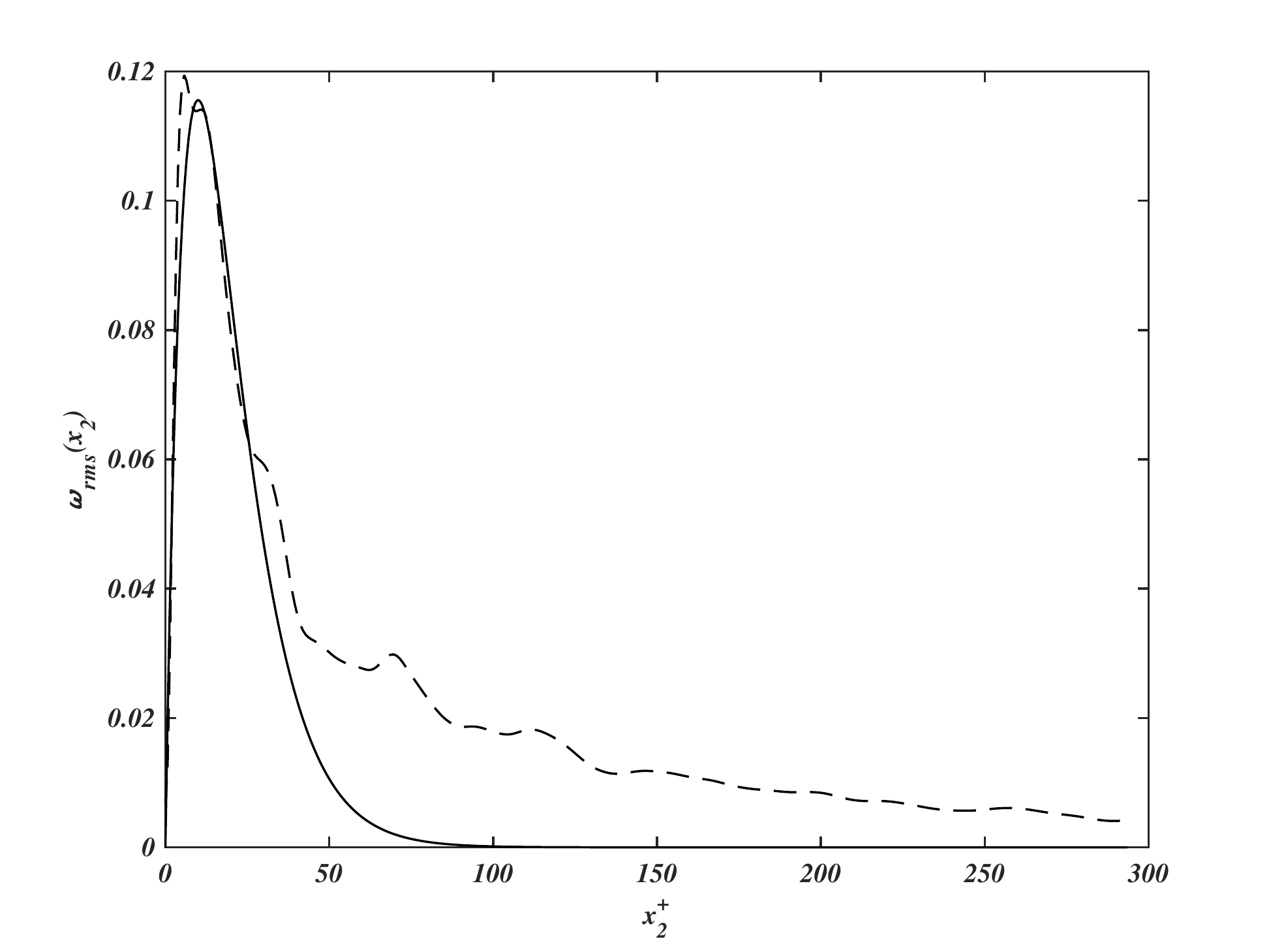}
\caption{\label{fig:Vorticity} A comparison of the rms vorticity as a function of distance from the wall, with the model defined by equation~\ref{eq:omegaRMSsqrd}. (dashed line DNS calculation, solid line model)}
\end{figure}

\begin{figure}
\centering
\includegraphics[width=0.8\linewidth]{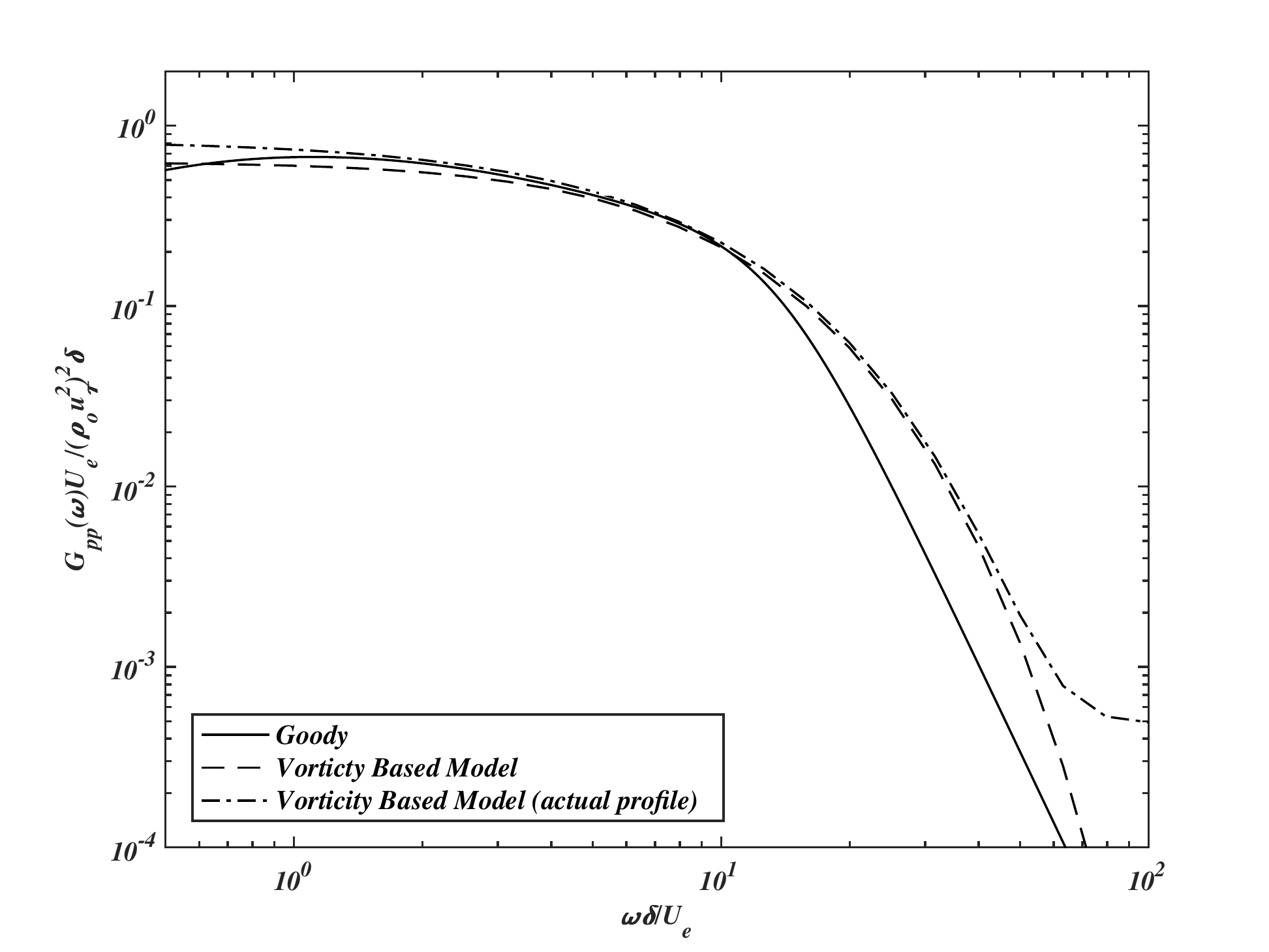}
	\caption{\label{fig:Goody} The predicted surface pressure spectrum compared to the Goody model. Solid line is the Goody model, the dashed line is the vorticity based model, and the dashed dotted line is the vorticity model with the mean square vorticity distribution obtained from the DNS.}
\end{figure}

\section{Discussion}
\label{sec:discussion}

In the previous section we have developed a modeling approach for the surface wavenumber spectra (equations~\ref{eq:phiPP2} and~\ref{eq:phiPP3}) and surface pressure spectra (equation~\ref{eq:SppNew}) in a turbulent boundary layer. Unlike previous methods, which are based on the solution to a Poisson's equation, the model presented here considers the solution to the OS equation in the viscous sublayer, and relates the surface pressure spectrum to the distribution of mean square vorticity as a function of height from the wall, the mean flow velocity profile, and a viscous length scale. The approach and modeling has been shown to be consistent with empirical models and DNS calculations of the flow, and so shows great promise for applications in more sophisticated flows that cannot be characterized by a zero pressure gradient parallel flow. For locally defined flows, the mean velocity profile and the mean square vorticity distribution can probably be obtained quite accurately from lower order models such as RANS, but this is beyond the scope of the present paper. The characteristic that is unlikely to be obtained from measurements or lower order models is the viscous length scale. However the DNS provides insight into this parameter since it resolves all the small scale turbulent structures.

\begin{figure}
\centering
\includegraphics[width=0.8\linewidth]{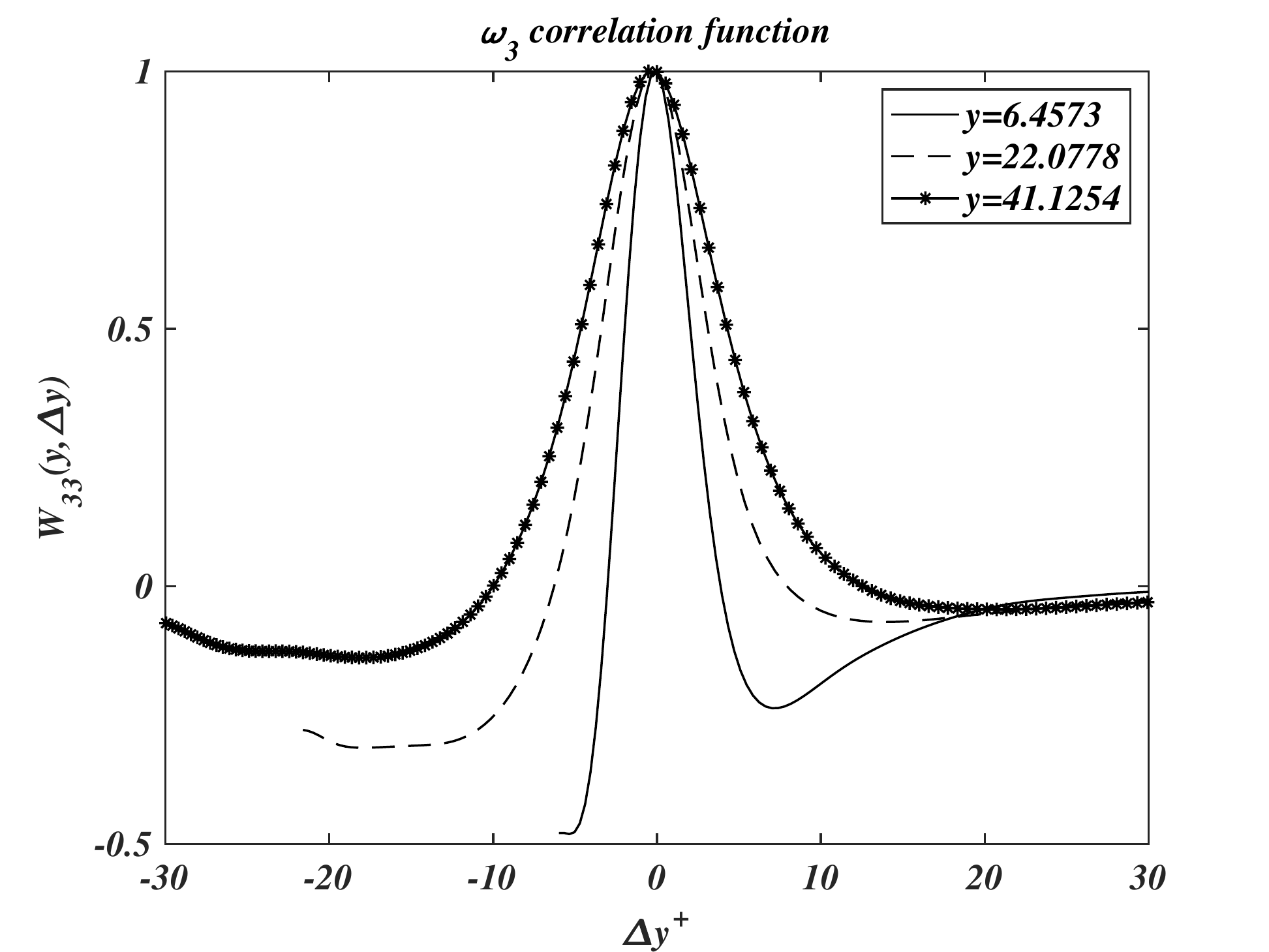}
	\caption{\label{fig:correlationWallNormal}The correlation functions of the spanwise vorticity as a function of displacement $\Delta y$ from a reference point at a height $y$ above the surface.}
\end{figure}

In Figure~\ref{fig:correlationWallNormal} we show the vorticity correlation functions as a function of distance normal to the wall for reference points that are at different heights from the surface. The correlation is not symmetric about the reference point as would be expected due to the presence of the wall. However the correlation becomes increasingly symmetric at larger distances from the wall. The most important conclusion from this figure is that the length scale is less than 10 wall units, at all heights. This scale is hard to resolve with RANS calculations, and implies that the vorticity distribution is only locally correlated. 

\begin{figure}
  \centering
\begin{subfigure}{0.8\textwidth}
  \centering
  \includegraphics[width=\linewidth]{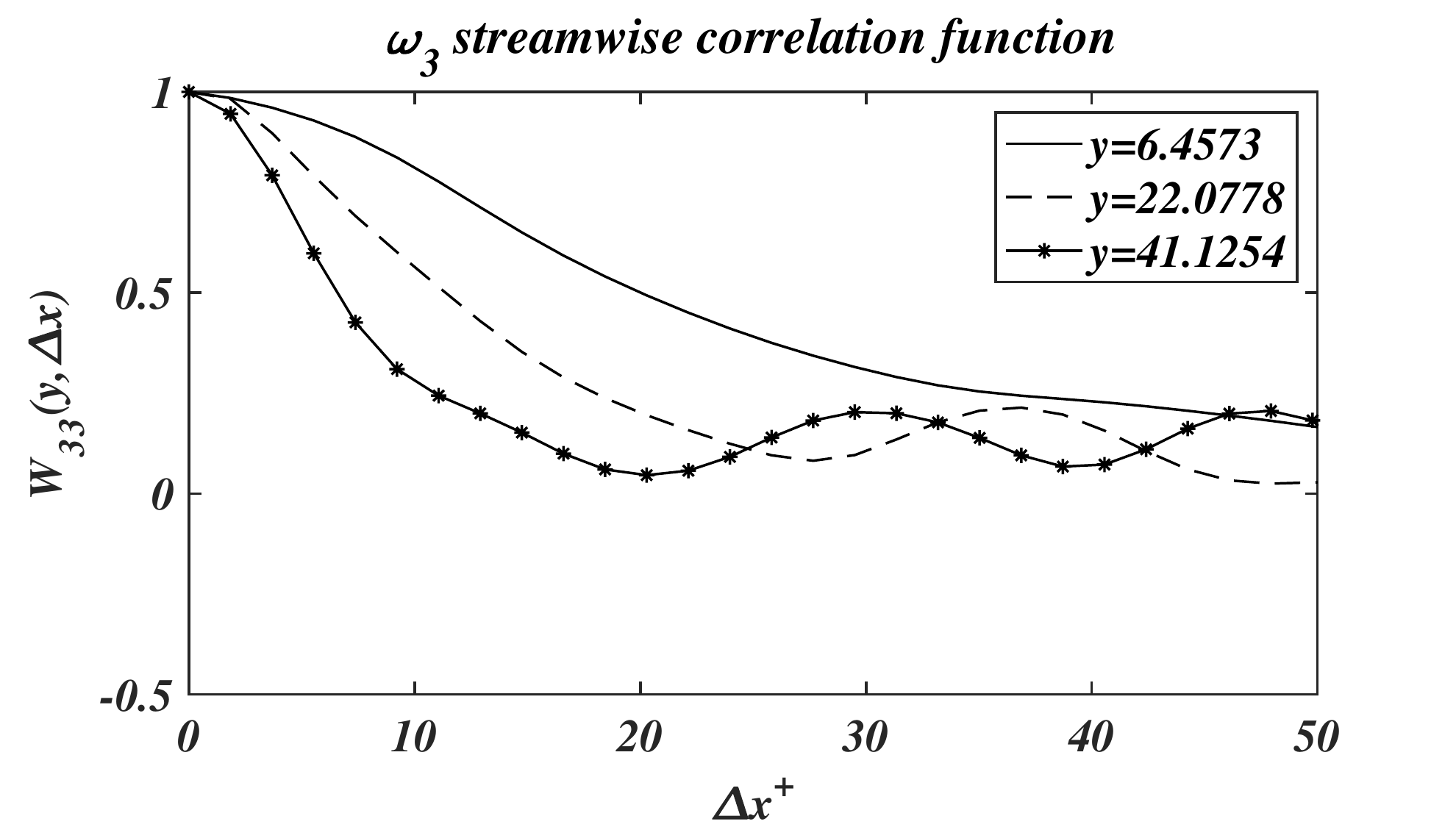}
  \caption{}
  \label{fig:sub-first}
\end{subfigure}
\begin{subfigure}{0.8\textwidth}
  \centering
  \includegraphics[width=\linewidth]{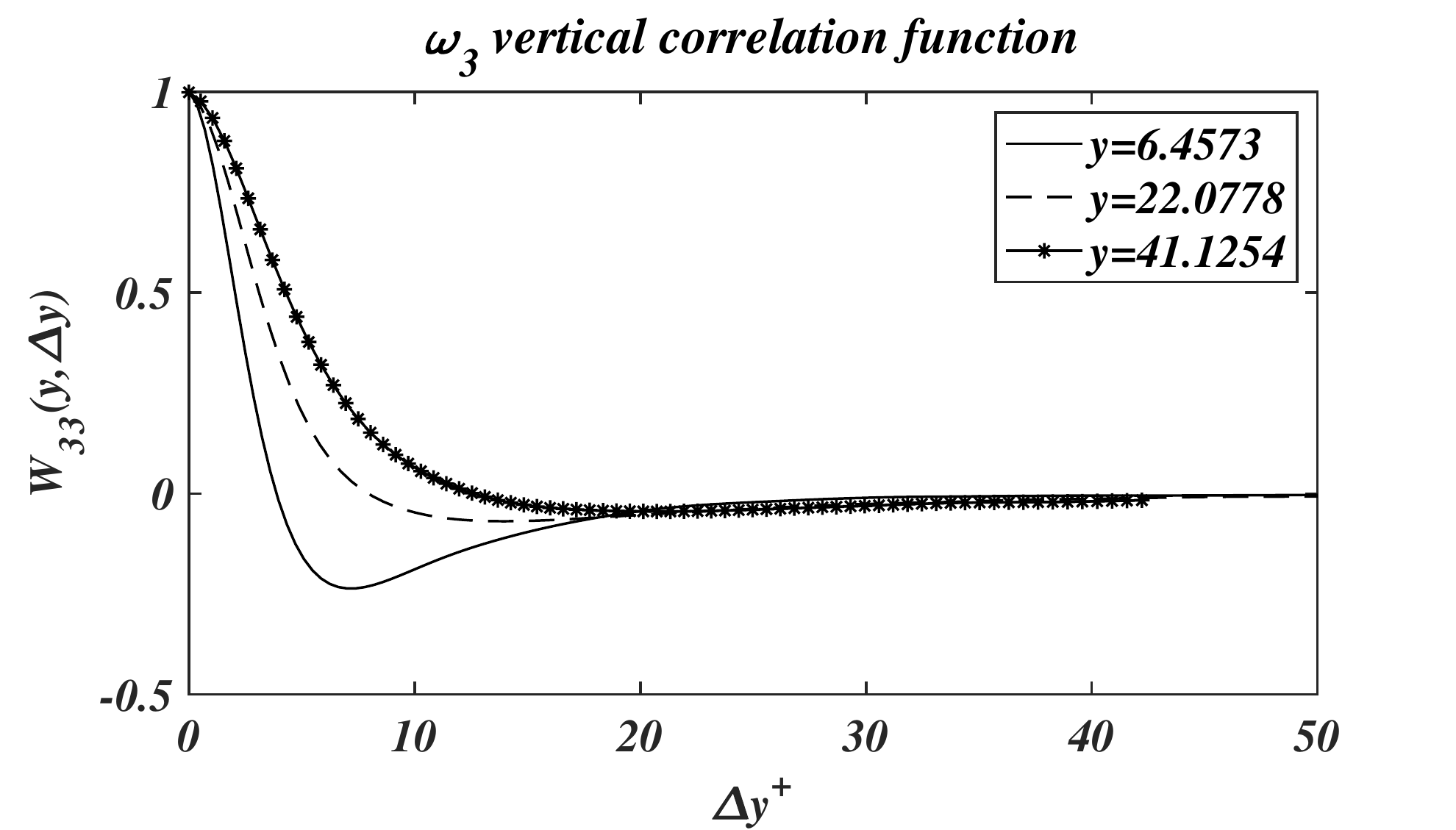}
  \caption{}
  \label{fig:sub-second}
\end{subfigure}
\begin{subfigure}{0.8\textwidth}
  \centering
  \includegraphics[width=\linewidth]{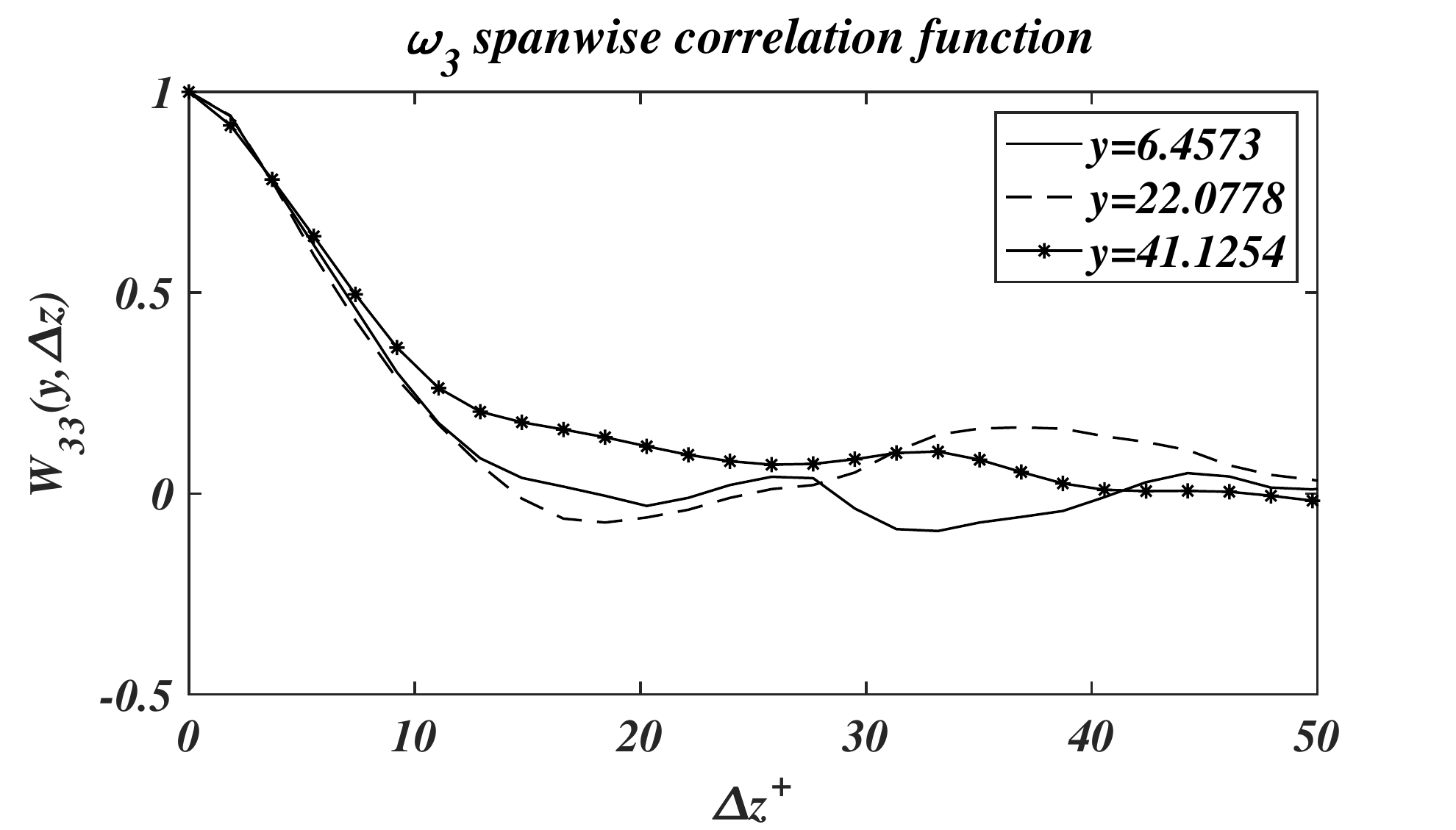}
  \caption{}
  \label{fig:sub-third}
\end{subfigure}
	\caption{\label{fig:correlStreamwiseSpanwise}The correlation functions of the spanwise vorticity for reference heights $y$ and displacements in (a) the streamwise direction, (b) the direction normal to the wall and (c) the spanwise direction.}
\end{figure}

The same conclusion can be drawn regarding the streamwise and spanwise vorticity as shown in Figure~\ref{fig:correlStreamwiseSpanwise}. Only the one sided correlation functions have been shown so that the isotropy of the flow can be assessed. These results are for the spanwise component of the vorticity and so do not include the effect of the streamwise vorticity that may be indicative of vortex streaks. However the streamwise vortex streaks are not strongly coupled with the aeroacoustics of these types of flow, such as trailing edge noise, and will only contribute at much lower frequencies than the spanwise vorticity. Not withstanding the importance of vortex streaks to the overall features of the flow, we see from Figure~\ref{fig:correlStreamwiseSpanwise} that the spanwise vorticity correlation functions are remarkably isotropic, and that their length scales are small. Close to the wall the streamwise length scale is significantly larger than the others (figure~\ref{fig:sub-first}, solid line), but as the reference point moves away from the wall to 22 wall units the lengthscales are very similar, and approximately equal to the 11 wall units needed to fit the Goody spectrum. It is very unlikely that such a small length scale could be measured accurately, which implies that to fully understand these flows high resolution numerical methods need to be used.

\section{Conclusion}
\label{sec:conclusion}

In this paper we have discussed the generation of surface pressure fluctuations below a turbulent boundary layer. It has been shown that the transmission of pressure through a shear flow is not fully described by a Poisson's equation, and to fully account for the effect of a mean shear flow the full set of the Navier Stokes equations must be solved. We have addressed this by considering the OS equation following the approach introduced by \citet{Landahl1967} and \citet{Chase1991}. By working in the wavenumber domain, it was shown how the flow variables can be specified in terms of an effective source term that depends on a set of source terms that are only dependent on non-linear shear stress fluctuations in the flow. Then by separating the OS equation into two coupled equations it was shown how the vorticity in the flow could be used as an effective source term for the surface pressure fluctuations if the vorticity associated with the wall shear stress was subtracted out. By considering a DNS calculation of a channel flow it was found that the influence of the near wall shear stress extended less than ten wall units from the surface, and the outer vorticity could be used to closely match the surface pressure wavenumber spectrum at a fixed instance in time over a wide range of wavenumbers.

One of the ongoing debates in the field of Aeroacoustics is the definition of the best `source term' that generates aerodynamic sound. Lighthill, Lilley, Howe and Goldstein have provided alternative formulations for the sound generation by flow, but all these formulations are limited by the complexity of the source terms and the difficulties in modeling them. In this paper we have introduced yet another approach, which is limited to incompressible flows and so will only apply in those situations where the sound is generated by an edge condition or surface discontinuities. We have shown that the surface pressure can be directly related to the vorticity in the flow outside the viscous sublayer. The formulation relies on the Biot Savart Law to relate the surface pressure to the vorticity, and the only approximation relates to the separation of the vorticity in the viscous sublayer that is required to match the no slip boundary condition. The key advantage of this formulation is that the vorticity in the flow can be modeled, whereas the non-linear source terms in the other acoustic analogies can be hard to measure or estimate. The vorticity source term will depend on the upstream boundary conditions, the vortex stretching, the mean shear lift up mechanism and the non linear vortex break up mechanism. Each of these can be studied separately, but, as we have shown, their combined effect is the driver of the surface pressure perturbation.

By using an approximation to the vorticity wavenumber spectrum we have derived a formula for the evaluation of the surface pressure spectrum, which gives a good fit to the measured surface pressure spectrum specified by \cite{Goody2004}. This is based on the assumption that the turbulence is locally convected by the mean flow and that limits the eddy convection speed to less than the free stream velocity, or the maximum flow speed in the boundary layer. This model is quite restrictive because the primary interest in surface pressure wavenumber spectra \citep{Graham1997} is at phase speeds that exceed the mean flow speeds in the boundary layer. The full solution, given by equation~\ref{eq:phiPP}, will include these phase speeds but its evaluation requires a more sophisticated model of the vorticity distribution. 

\section*{Declaration of Interests}
The authors report no conflict of interest.

\section*{Acknowledgements}
This work is supported by ONR grant number N00014-20-1-2774. The author would like to thank Dr. Ki-Han Kim, Dr. Yin Lu Young and Dr. John Muench for their support of this project. The work follows on from a study supported by NSF Grant CBET-1802915. Computational resources were provided by the ONR under HPC grant ONRDC48942624. In addition the author would like to thank Dr. William Devenport, Nandita Hari, Matt Szoke, Alexander Gonzalez, Ignacio Jimenez, Frank Ballestieri, Dr. William Blake and Dr. Jason Anderson for valuable discussion on this topic and inspiring many of the ideas presented.


\bibliographystyle{jfm}
\bibliography{bibFile}

\end{document}